\documentclass[12pt]{article}
\usepackage{graphicx}
\usepackage{multirow}

\newcommand\pubnumber{CERN-TH-2018-006}
\newcommand\pubdate{\today}

\def\institute{Theoretical Physics Department, CERN, Geneva, Switzerland}
\def\support{\footnote{Work supported by the Korean Research Foundation (KRF) through the CERN-Korea Fellowship program.}}

\def\Title#1{\begin{center} {\Large #1 } \end{center}}
\def\Author#1{\begin{center}{ \sc #1} \end{center}}
\def\Address#1{\begin{center}{ \it #1} \end{center}}

\newcommand\pubblock{\rightline{\begin{tabular}{l} \pubnumber\\
         \pubdate  \end{tabular}}}
\newenvironment{Abstract}{\begin{quotation}  }{\end{quotation}}
\newenvironment{Presented}{\begin{quotation} \begin{center} 
             PRESENTED AT\end{center}\bigskip 
      \begin{center}\begin{large}}{\end{large}\end{center} \end{quotation}}
\def\Acknowledgements{\bigskip  \bigskip \begin{center} \begin{large}
             \bf ACKNOWLEDGEMENTS \end{large}\end{center}}

\def\beq{\begin{equation}}
\def\eeq#1{\label{#1}\end{equation}}
\def\eeqn{\end{equation}}

\def\beqa{\begin{eqnarray}}
\def\eeqa#1{\label{#1}\end{eqnarray}}
\def\eeqan{\end{eqnarray}}

\let\bar=\overbar

\def\Dslash{\not{\hbox{\kern-4pt $D$}}}
\def\dslash{\not{\hbox{\kern-2pt $\del$}}}

\def\msb{{\bar{\ssstyle M \kern -1pt S}}}

\newcommand{\mellinO}{\mathcal{M}_{\mathcal{O}}}
\newcommand{\DeltaMO}{\Delta_{m_{t}}^{(\mellinO)}}

\newcommand{\text}[1]{\textrm{#1}}

\begin{document}
\begin{titlepage}
\pubblock

\vfill
\Title{Understanding the Impact of Pythia Hadronization in Top Quark Mass Determinations at the LHC}
\vfill
\Author{ Doojin Kim\support}
\Address{\institute}
\vfill
\begin{Abstract}
Most of the methods to measure the top quark mass suffer from the jet energy scaling in achieving better precision. As a way to circumvent this issue, the study of $B$-hadron observables is motivated. While they do not involve such an issue, understanding underlying hadronization models is a key to achieve $\sim0.5\%$ precision or better. In this presentation, we discuss the impact of the hadronization model parameters -- for example, implemented in Pythia 8 -- on precision measurements of the top quark mass through $B$-hadron observables. We study the sensitivity of the top quark mass to relevant hadronization and showering parameters, followed by a discussion on observables to be used for constraining the hadronization and showering parameters.
\end{Abstract}
\vfill
\begin{Presented}
$10^{th}$ International Workshop on Top Quark Physics\\
Braga, Portugal,  September 17--22, 2017
\end{Presented}
\vfill
\end{titlepage}
\def\thefootnote{\fnsymbol{footnote}}
\setcounter{footnote}{0}

\section{Introduction}

Precision top quark mass measurement is of paramount importance in both the Standard Model (SM) and new physics models beyond the SM; for example, SM vacuum stability. 
Most of the standard techniques for determining the top quark mass, $m_t$, require the reconstruction of bottom quark-induced jets, so a major source of uncertainty in those methods is the $b$-jet energy scaling (bJES)~\cite{ATLAS:2014wva}.
To get around this issue, we propose here an $m_t$ measurement method using various experimental observables formed with $B$ hadrons~\cite{Corcella:2017rpt}.
At the cost of avoiding bJES, however, it is crucial to understand the transformation from $b$ quark to $B$ hadron, which is quite challenging as it is governed by non-perturbative QCD.  
To address this challenge, we investigate the impact of the hadronization model parameters, especially implemented in Pythia 8~\cite{Sjostrand:2014zea}, on the $m_t$ determination via $B$-hadron observables.
We begin with a brief explanation about the strategy that we employ in this article. We then present our main results, i.e., $m_t$ sensitivity to relevant hadronization and showering parameters, followed by a discussion on variables to be used for constraining those parameters.

\section{Strategy}

For our study, we generate the leading-order, fully leptonic events of top quark pairs produced at the LHC-13 TeV with Pythia 8.2~\cite{Sjostrand:2014zea}. Jets are reconstructed by the anti-$k_t$ algorithm of $R=0.5$. As selection criteria, we then require two hardest final-state jets and two leptons to obey the following cuts:
\beq
p_{T,j}>30 \textrm{ GeV},\, |\eta_j|<2.4,\,\,\,p_{T,\ell}>20 \textrm{ GeV},\, |\eta_\ell|<2.4,
\eeqn
where $j$ and $\ell$ denote jets and leptons (electrons and/or muons), respectively. Simulations are performed with 21 input $m_t$ values, from 163 GeV to 183 GeV by an interval of 0.5 GeV.
For each mass point, three hadronization parameters relevant to heavy flavors and the formation of $B$ hadrons, which are tabulated in the last three rows in Table~\ref{tab:varied}, are varied with the ranges appearing in the third column of Table~\ref{tab:varied}.
Since hadronization models are specific to the showering environment to which they are attached, we also vary a few relevant showering parameters whose corresponding Pythia switches and ranges are summarized again in Table~\ref{tab:varied}.


\begin{table}[t]
\begin{centering}
\small{
	\begin{tabular}{c c c}
	\hline 
	 & Pythia 8 parameter & Range (central value) \tabularnewline
	\hline  
	$p_{T,\min}$ & \textsc{TimeShower:pTmin } & $0.25-1.00$ (0.5) GeV \tabularnewline
	$\alpha_{s, \textrm{{\scriptsize FSR}}}$ & \textsc{TimeShower:alphaSvalue } & $0.1092 - 0.1638$ (0.1365) \tabularnewline 
	recoil  & \textsc{TimeShower:recoilToColoured} & \emph{on} and \emph{off}(\emph{on})\tabularnewline 
	$b$ quark mass & \textsc{5:m0} & $3.8-5.8$ (4.8) GeV \tabularnewline 
	Bowler's $r_{B}$  & \textsc{StringZ:rFactB } & $0.713-0.813$ (0.855) \tabularnewline 
	string model $a$ & \textsc{StringZ:aNonstandardB } & $0.54-0.82$ (0.68) \tabularnewline 
	string model $b$ & \textsc{StringZ:bNonstandardB } & $0.78-1.18$ (0.98) \tabularnewline
	\hline 
	\end{tabular}
	}
\par\end{centering}
\caption{\label{tab:varied}
A summary of relevant Pythia 8 parameters and their variation ranges.}
\end{table}

Several observables are adopted to study $m_t$ sensitivity to various Pythia parameters, which are listed below.
\begin{itemize} \itemsep1pt \parskip0pt \parsep0pt
\item $E_B$: $B$-hadron energy~\cite{Agashe:2016bok},
\item $p_{T,B}$: $B$-hadron transverse momentum,
\item $E_{B}+E_{\bar B}$ and $p_{T,B}+p_{T,\bar B}$: sum of the energies
  and transverse momenta of $B$ and $\bar{B}$,
\item $m_{B\ell}$: invariant mass constructed by the $B$-hadron and one
lepton from $W$ decay,
\item $m_{T2}$~\cite{Lester:1999tx} and $m_{T2,\perp}$~\cite{Matchev:2009ad} of the $B$ and the $B\ell$ subsystems,
\item $m_{BB\ell\ell}$, the total mass of the system formed by the two $B$ hadrons and the two leptons in an event.
\end{itemize}
Note that the $m_{B\ell}$ and $m_{T2}$ (and $m_{T2,\perp}$) of the $B\ell$ subsystem involve the combinatorial ambiguity, and a subtlety arises in defining the missing transverse momentum for $m_{T2}$ and $m_{T2,\perp}$ variables.
So, we refer to Ref.~\cite{Corcella:2017rpt} for detailed prescriptions and definitions.

For constraining the parameters listed in Table~\ref{tab:varied}, we take a number of calibration variables which have no/little sensitivity to the input top quark mass but decent sensitivities to Pythia parameters in $t\bar{t}$ events.
\begin{itemize} \itemsep1pt \parskip0pt \parsep0pt
\item $p_{T,B}/p_{T,j_{b}}$, 
\item $\rho(r)=\frac{1}{\Delta r}\frac{1}{E_{j}}\sum_{\textrm{track}}E(\textrm{track})\theta\left(\left|r-\Delta R_{j,\textrm{track}}\right|<\delta r\right)$:  the radial jet-energy density which were defined and measured in~\cite{Aad:2013fba},
\item $\chi_{B}=2E_{B}/X_{B}$ with $X$ possibly being
   $m_{j_{b}{j}_{\bar{b}}}$, $\sqrt{s_{\min}}$~\cite{Konar:2010ma},
   \ $\left|p_{T,j_{b}}\right|+\left|p_{T,{j}_{\bar{b}}}\right|$ and
   $E_{j_b}+E_{j_{\bar b}}$,
\item $\Delta\phi(j_{b}j_{\bar{b}})$, $\Delta\phi(B\bar B)$, $|\Delta\phi(j_{b}j_{\bar{b}})-\Delta\phi(B\bar B)|$, $\Delta R(j_{b}j_{\bar{b}})$, $\Delta R(B\bar B)$,
and $|\Delta R(j_{b}j_{\bar{b}})-\Delta R(B\bar B)|$,
\item $E_{B}/E_{\ell}$, $E_{B}/(E_{\ell}+E_{\bar{\ell}})$.
\end{itemize}
Since these variables are sensitive to different kinematical and/or dynamical configurations in association with the Pythia parameters, we anticipate that they contain independent information to constrain those parameters~\cite{Corcella:2017rpt}.  

Finally, we express our sensitivity measure of a given observable $\mathcal{O}$ to a generic Monte Carlo parameter $\theta$ in terms of logarithmic derivatives:
\beq
\Delta_{\theta}^{(O)}\equiv\frac{\bar{\theta}}{\bar{O}}\frac{\partial O}{\partial\theta},\label{eq:DeltaGenericDef}
\eeqn
where the barred quantities $\bar{\mathcal{O}}$ and $\bar{\theta}$ indicate corresponding average values with respect to a given range of $\theta$ shown in Table~\ref{tab:varied}.

\section{Results and Discussions}

We first report our results on the $m_t$ measurement with $B$-hadron observables. For each set of Monte Carlo parameters and input $m_t$, we conduct the first Mellin moments (henceforth denoted as $\mathcal{M}_{\mathcal{O}}$) analyses of the observables and the shape (e.g., kinematic endpoint) analyses of some observables. For the endpoints, we employ a second-order polynomial, $y(x)=c_1(x-x_m)^2+c_2(x-x_m)$, with $x_m$ identified as the endpoint and $c_{1,2}$ being other fit parameters irrelevant to our analysis.

\begin{table}[t]
\begin{centering}
\hspace*{-0.4cm}\small{
\begin{tabular}{c|c|c|c c c c c c c}
\hline 
\multirow{2}{*}{$\mathcal{O}$} & Range & \multirow{2}{*}{$\DeltaMO$} & \multicolumn{7}{c}{$\Delta_{\theta}^{(m_{t})}$}\tabularnewline
\cline{4-10} 
 & [GeV] &  & $\alpha_{s,\text{{\tiny FSR}}}$ & $m_{b}$ & $p_{T,\min}$ & $a$ &  $b$ & $r_{B}$  & recoil\tabularnewline
\hline  
$E_{B}$ & 28-110 & 0.92 & -0.52 & -0.21 & 0.057 & -0.02 & 0.06 & -0.10 & -0.022\tabularnewline 
$p_{T,B}$ & 24-72 & 0.92 & -0.54 & -0.21 & 0.056 & -0.03& 0.07 & -0.09 & -0.023\tabularnewline 
$m_{B\ell}$ & 47-125 & 1.30 & -0.241 & -0.072 & 0.022 & -0.007 & 0.023 & -0.02 & -0.008\tabularnewline  
$E_{B}+E_{B}$ & 83-244 & 0.92 & -0.50 & -0.21 & 0.056 & -0.02 & 0.07 & -0.08 & -0.020\tabularnewline
$m_{BB\ell\ell}$ & 172-329 & 0.96 & -0.25 & -0.10 & 0.028 & -0.01 & 0.026 & -0.03 & -0.008\tabularnewline 
$m_{T2,B\ell}$  & 73-148 & 0.95 & -0.27 & -0.09 & 0.029 & -0.009 & 0.03 & -0.03 & -0.010\tabularnewline
\hline 
$m_{B\ell}$ & 127-150 &1.26& 0.017 & 0.003 & -0.006 & -0.008 &  0.008 & -0.016 &  -0.00042 \tabularnewline
$m_{T2,B\ell}$ & 150-170 & 0.98 & -0.01 & -0.023 & 0.007 & -0.006 & 0.010 & -0.011 & -0.0002 \tabularnewline
\hline 
\end{tabular}
}
\par\end{centering}
\vspace{-0.3cm}
\caption{\label{tab:Sensitivity-Mellin-Mass} A partial list of $m_t$ determination results. The last two rows are for the shape analysis, while the others are for the Mellin moment analysis.}
\vspace{0.3cm}
\begin{centering}
\hspace*{-2.5cm}\small{
\begin{tabular}{c|c|c|c c c c c c c}
\hline 
\multirow{2}{*}{$\mathcal{O}$} & \multirow{2}{*}{Range} & \multirow{2}{*}{$\DeltaMO$} & \multicolumn{7}{c}{$\Delta_{\theta}^{(\mathcal{M}_{\mathcal{O}})}$}\tabularnewline
\cline{4-10} 
 & &  & $\alpha_{s,\text{{\tiny FSR}}}$ & $m_{b}$ & $p_{T,\min}$ & $a$ &  $b$ & $r_{B}$  & recoil\tabularnewline
\hline  
$\rho(r)$ & 0-0.44 & -0.007 & 0.78 & 0.204 & -0.1286 & 0.029 & -0.043 & 0.056 & 0.020\tabularnewline 
$p_{T,B}/p_{T,j_b}$ & 0.6-0.998 & -0.053 & -0.220 & -0.1397 & 0.0353 & -0.0187& 0.0451 & -0.0518 & -0.0108\tabularnewline 
$|\Delta R(B\bar B)-\Delta R (j_b j_{\bar b})|$ & 0-0.0992 & 0.10 & 0.920 & 0.079 & -0.075 & -0.000 & 0.005 & -0.00 & 0.418\tabularnewline  
\hline 
\end{tabular}
}
\par\end{centering}
\vspace{-0.3cm}
\caption{\label{tab:calib} A partial list of calibration observable results.}
\end{table}   

A partial list of our $m_t$ determination results are shown in Table~\ref{tab:Sensitivity-Mellin-Mass} (see Ref.~\cite{Corcella:2017rpt} for the full results). 
While looking at numbers under the column of $\Delta_{\theta}^{(m_{t})}$, we observe that the Mellin moment results are most sensitive to $\alpha_{s,\text{{\scriptsize FSR}}}$. Note that the endpoint results are less sensitive to the Pythia parameters because they are minimally affected by underlying dynamics due to their kinematic nature. However, they might suffer from securing enough events even with the high-luminosity LHC because events usually populate less near the endpoints. 
In any case, it is crucial and highly motivated to improve knowledge on the relevant Monte Carlo parameters in order to achieve better precision in the $m_t$ measurement.

For calibration variables listed earlier, we again perform a Mellin moment analysis. It turns out that Mellin moment observables allow us to constrain only one linearly independent combination of Pythia parameters with precision similar to that in the measurement of observables. More quantitatively, the relevant singular values of our final sensitivity matrix $(\Delta_{\theta}^{(\mathcal{M}_{\mathcal{O}})})_{ij}$ are 1.7, 0.6, 0.048, 0.0075, 0.005, 0.0033, and 0.0014. A few representative entries of the matrix are reported in Table~\ref{tab:calib}, while we refer to Ref~\cite{Corcella:2017rpt} for the full set of results. Indeed, our correlation study among the obeservables suggests that most of them contain not much orthogonal information. 

In light of this challenge, we propose to utilize directly the bin counts of a subset of the calibration variables in suitably selected ranges. Our choice of those variables are $\rho(r)$, $\chi_B(E_{j_b}+E_{j_{\bar b}})$, $E_B/E_\ell$, $|\Delta R(B\bar B)-\Delta R (j_b j_{\bar b})|$, $p_{T,B}/p_{T,j_b}$, and $m_{BB}/m_{j_b j_{\bar b}}$ which showed the greatest sensitivity and the most distinctive dependence on linearly independent combinations of Pythia parameters in the Mellin moment analysis. Conducting a global analysis on the above 6 observables, we find that the singular values of the associated sensitivity matrix are 15.0, 4.2, 0.75, 0.42, 0.27, 0.16, and 0.13. This implies that input observables with 1 \% precision should give $\sim10$ \% precision even on the most loosely constrained parameter.

In summary, we discussed the $m_t$ determination with $B$-hadron observables and the potential impact of Pythia hadronization and showering parameters on them. We found that they are most sensitive to $\alpha_{s, \textrm{{\scriptsize FSR}}}$, so a careful ``tune'' with the proposed calibration strategy will reduce the theoretical uncertainty of $m_t$ in $B$-hadron observables. More quantitatively, $\alpha_{s, \textrm{{\scriptsize FSR}}}$ should be constrained at $1-2$ \% level while the others should be constrained at $10-20$ \% level to achieve $\sim0.5$ \%  precision in $m_t$.

\Acknowledgements
This work is supported by the Korean Research Foundation (KRF) through the CERN-Korea Fellowship program.

\end{document}